\documentclass[epj,twocolumn]{webofc}
\usepackage[varg]{txfonts}   
\begin{document}
\title{Model-Stable Universality of the Air Shower Electromagnetic Component:}
%
%
\subtitle{an Approach to Solving the Mass Composition Problem}

\author{\firstname{R.} \lastname{Raikin}\thanks{\email{raikin@theory.asu.ru}},
        \firstname{T.} \lastname{Serebryakova},
        \firstname{A.} \lastname{Lagutin},
        \firstname{N.} \lastname{Volkov}\inst{1}
}

\institute{Altai State University, 61 Lenin Ave., Barnaul, 656049, Russia}

\abstract{%
On the basis of the scaling approach and CORSIKA simulations data the radial scale factor of lateral distribution of electrons in extensive air showers is confirmed as potentially effective primary mass estimator, and its sensitivity to hadronic interaction model is investigated. It is shown that improved composition results both on average and event-by-event basis can be achieved taking into account the universality property of air shower development expressed by the relation between radial scale factor and longitudinal age parameter. The enhancements of such a theoretically motivated tool for the unbiased cosmic ray composition deduction in a wide primary energy range from (multi-)hybrid air shower measurements of nearest future are discussed.}
\maketitle
\section{Introduction}

Uncertainties in hadronic physics and chemical composition are two basic obstacles for understanding the origin of very high energy cosmic rays. Physical interpretation of features of cosmic ray energy spectrum in terms of sources and propagation properties relies on the assumed mass composition while its robust estimation from extensive air shower (EAS) observables interferes with their sensitivity to nuclei interaction model at energies not accessible with the accelerators.

At present stage numerous methods and techniques are implemented to infer the mass composition of cosmic rays~(see e.g.~\cite{KampertUnger}). They include the analysis of mean values, fluctuations, correlations and even the particular features of distributions of different EAS observables such as depth of shower maximum, muon production depths, total number of electrons and muons at the observation level and local densities far from the shower axis, as well as particles arrival time distributions and spatial distribution of EAS radio signal. However, despite considerable efforts that have been made in recent years, the composition and its variation with energy remain quite uncertain in almost entire primary energy range of the cosmic rays available for EAS studies~\cite{Haungs, Haungs2, KT, buitink2016}.

In this paper we examine the efficiency of the formalism describing lateral distribution function (LDF) of EAS electrons as scale-invariant and its dependence on shower longitudinal development stage, that reflects EAS universality properties, for reducing the uncertainties in current analysis and for the improvement of estimation either the mean mass composition at a certain energy or primary particle type in case of individual showers.

In the following section we give a brief overview of the scaling approach for LDF, proposed in~\cite{Lagutin:1998fx, Durban,Raikin:predictions,RaikinJPG} and the outline for the following simulations data analysis. In section 3 we present the results of CORSIKA simulations of EAS carried out with EPOS LHC and SIBYLL nuclei interactions models: distributions of depth of shower maximum $X_{\rm max}$ and radial scale factor $R_0$ for vertical showers initiated by protons and iron nuclei, correlations between $R_0$ and longitudinal age parameter $s$. Section 4 contains discussion and final conclusions.

\section{Radial scaling of electron component and universality in EAS development}

One of the key EAS quantities necessary for basic shower parameters reconstruction is the lateral distribution of charged
particles at fixed observation depth $X$. The exact form of LDF has been debatable for decades. The majority of
analytical parametrizations of LDF of different EAS components is traditionally based on the well known Nishimura-Kamata-Greizen
(NKG) function~\cite{Greizen1960} originally obtained for electromagnetic cascade showers:
\begin{eqnarray}
\rho(r;E,s)=\frac{N(E,s)}{r_0^2}\frac{\Gamma(4.5-s)}{2\pi\Gamma(s)\Gamma(4.5-2s)}\times\nonumber\\
\times\left(\frac{r}{r_0}\right)^{s-2}\left(1+\frac{r}{r_0}\right)^{s-4.5}.
\label{NKG}
\end{eqnarray}
Here $\rho(r;E,s)$ is local particle density at radial distance
$r$ from the core position in shower with primary energy $E$ and
the longitudinal age parameter~$s$, $N(E,s)$~--- total number of particles at
the observation depth (shower size), $r_0$~--- shower scale radius, which does
not depend on primary particle type and energy (originally~--- the
Moliere unit $r_M$). Various modifications of NKG form, such as
introducing different fixed scale factors, lateral ($s_\perp$) or local ($s(r)$) age parameters and also generalizations of the
function by using third power-law term were suggested. A comprehensive review is beyond the scope of this paper (some discussions could be found in e.g.~\cite{NKGr1,NKGr2,NKGr3}).

A different theoretically motivated approach, so-called scaling formalism, was proposed in our
papers~\cite{Lagutin:1998fx, Durban,Raikin:predictions,RaikinJPG} for the lateral distribution of electrons in both electromagnetic and hadronic showers:
\begin{equation}
\rho(r;E,X)=\frac{N(E,X)}{R_0^2(E,X)}\,F\left(\displaystyle\frac{r}{R_0(E,X)}\right).
\label{scaling1}
\end{equation}
Here the radial scale factor $R_0$, in contrast to commonly used $r_M$, depends on primary particle type, shower age and (in case of extensive air showers) properties of hadronic interactions. Function $F(x)$ is the common scaling part of LDF.
According to our calculations based on semi-analytical approach~\cite{Lagutin:1998fx, Durban, RaikinJPG}, factor $R_0$ is equal to the root mean square radius of electron component $R_{\rm ms}$, which is defined in a standard way as
\begin{equation}
R_\text{ms}(E,X)=\left(\displaystyle\frac{2\pi}{N(E,X)}\int_0^\infty r^2 \rho(r;E,X)rdr\right)^{1/2},
\end{equation}
and the following expression suggested in~\cite{Raikin:predictions} for $F(x)$ could be used
\begin{equation}
F(x)=Cx^{-\alpha}(1+x)^{-(\beta-\alpha)}(1+(x/10)^\gamma)^{-\delta},
\label{lrf}
\end{equation}
$C=0.28$, $\alpha=1.2$, $\beta=4.53$, $\gamma=2.0$, $\delta=0.6$.

On the basis of simulations using the simplified hadronic generator, it was also found~\cite{RaikinJPG,Raikin:2013IzvRan,Raikin:2013JPhys} that the relation between the lateral shape of the electron distribution and the longitudinal shower age can be expressed by the $R_0(s)$ functional dependence, that is a consequence of shower universality properties~\cite{Lipari}.

\section{Results}

Monte-Carlo simulations of EAS initiated by protons and iron nuclei in the energy range $10^{15}\div 10^{19}$~eV were performed using CORSIKA v.7.4100~\cite{CORSIKA} with EPOS LHC v.3400 and SIBYLL v.2.1 (FLUKA 2011.2c.2) hadronic interactions models. In order to get reliable data on electron LDF at very large distances from the shower core the thinning level and particle weight limit were set as $\varepsilon_{th}= 10^{-8}$ and $\omega=10^2$ respectively.

The analysis showed that for the averaged lateral distributions the whole set of calculated electron densities are reproduced by scaling representation~(\ref{scaling1}) within 20\% in relative discrepancy over the interval of scaling variable $x=r/R_0=(0.05\div 25)$, corresponding to the region of radial distances from $r\sim 10$~m to $r\sim (2\div 4)$~km. For smaller distances discrepancies become large. It does not allow to use the scaling approach for shower size evaluation from the experimentally measured local densities far from the axis. This limitation of the scaling approach was also noted in~\cite{Kalm1, Kalm2, Fomin}. Nevertheless, scaling formalism provides precise description of the shape of lateral distribution of electrons at radial distances, where local densities can be measured by ground detectors of large air shower arrays. It should be particularly emphasized that the dependence of LDF on hadronic interaction model is completely governed by variation of single parameter $R_0$.

Radial scale factors were evaluated for both average and individual showers as parameters of simulated LDFs fitting. Note, that to obtain the unbiased estimates of $R_0$ iterative procedure was implemented for discrimination of data at distances where scaling formalism is not valid and also for using fixed distances ranges with respect to scaling variable $x=r/R_0$ for showers of specific energy and primary particle type. For $F(x)$ expression~(\ref{lrf}) was used. We have checked out function~(\ref{lrf}) with a refined set of parameters $\alpha,~\beta,~\gamma,~\delta$ and other representations, including polynomial approximation for $x^2F(x)$ giving considerably better overall fit of simulated data. It was found, that the resulting $R_0$ values demonstrate only weak dependence on the explicit form of $F(x)$ choosing for fitting.

\begin{figure}
\centering
\includegraphics[width=0.475\textwidth,clip]{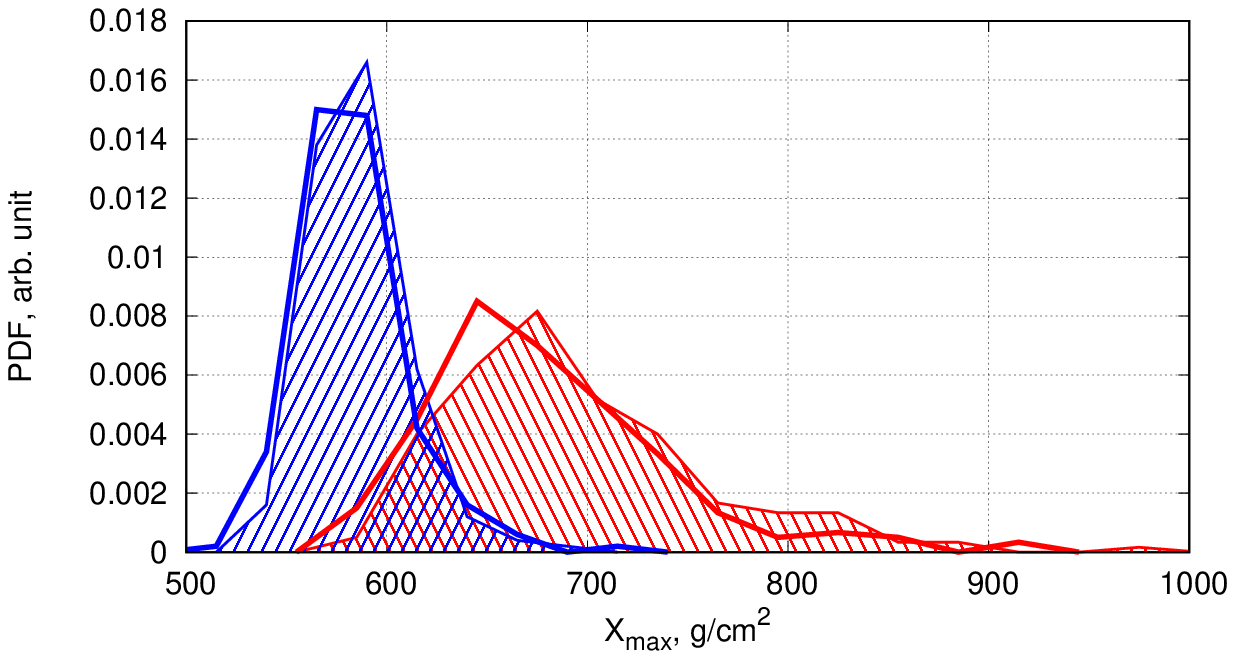}
\includegraphics[width=0.475\textwidth,clip]{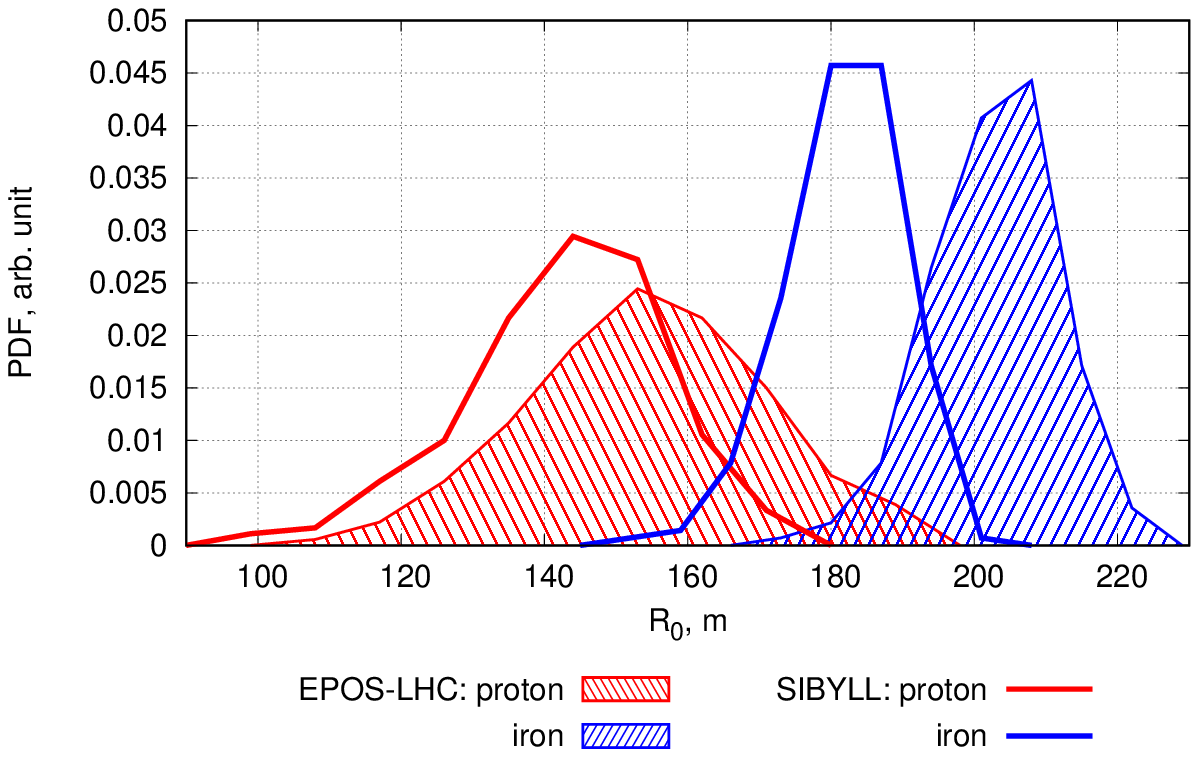}
\caption{Distributions of depth of shower maximum $X_{\rm max}$ (top) and radial scale factor $R_0$ (bottom) at sea level for $10^{17}$~eV vertical showers initiated by protons (red) and iron nuclei (blue).}
\label{fig-1}
\end{figure}

The distributions of depth of shower maximum $X_{\rm max}$ and radial scale factor $R_0$ at sea level for $10^{17}$~eV vertical showers initiated by protons and iron nuclei are shown in Figure~\ref{fig-1}. 200 simulated showers are included in each data set. Results for EPOS LHC and SIBYLL interaction models are shown with shaded and open distributions respectively. The figure clearly displays that $R_0$ estimated from simulations is more sensitive to both primary mass and interaction model in comparison with $X_{\rm max}$.

\begin{figure}
\centering
\includegraphics[width=0.475\textwidth,clip]{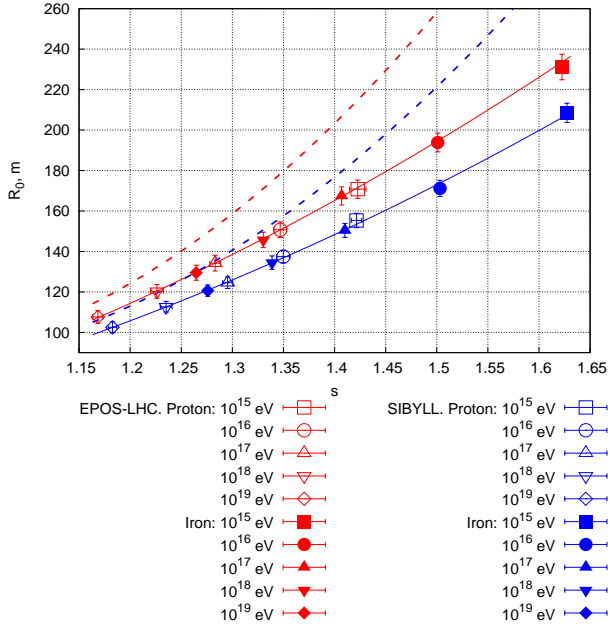}
\caption{Radial scale factor $R_0$ for average EAS of fixed energy as a function of shower age $s$ (see text for details).}
\label{fig-2}
\end{figure}

Figure~\ref{fig-2} shows the scale factor $R_0$ for average EAS of fixed energy at sea level versus the longitudinal shower age $s$, defined in terms of a classic cascade theory as $s(X, X_{\rm max})=3/(1+2X_{\rm max}/X)$. At the first stage electrons produced from the decay of low energy muons were eliminated from local particle densities. In this case electron LDF is a pure superposition of partial electromagnetic subshowers definitely possessing scaling and universality features. This data is shown by symbols and solid lines of red (EPOS LHC) and blue (SIBYLL) color. When the contribution of electrons from muon decays is taking into account the LDF at very large distances become narrower, which results in increased $R_0$. This effect wanes with energy. The corresponding data is presented by dashed lines in Figure~\ref{fig-2}. It is important that in both cases there is a functional dependence $R_0(s)$, i.e. one-valued relation between parameters of shower lateral ($R_0$) and longitudinal ($s$) development stage representing the air shower universality~\cite{Raikin:2013IzvRan, Raikin:2013JPhys}.

One can see from Figure~\ref{fig-2}, that the rate of decrease of $R_0$ with energy is almost equal for two interaction models used in calculations. Thus the $\partial R_0/\partial \log E$ value can be suggested~\cite{RaikinJPG, Raikin:2008zz, Raikin:2009zz, Raikin:2011IzvRan, Raikin:2011zz} as a model-independent measure of primary composition variations with energy similarly to the widely used elongation rate approach.

\begin{figure}
\centering	
\sidecaption
\includegraphics[width=0.475\textwidth,clip]{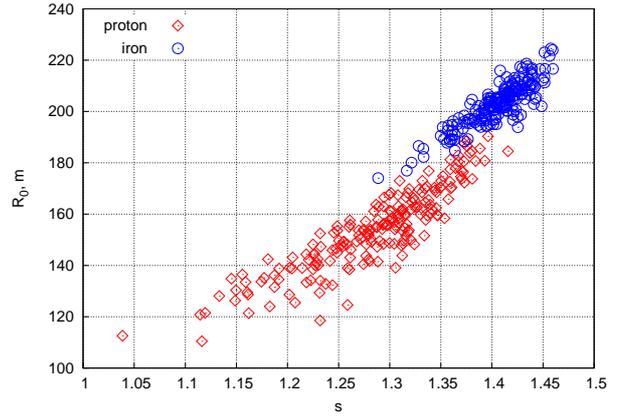}
\caption{$R_0$ vs $s$ at sea level scatter plot. Data for 200 simulated vertical showers initiated by protons (red diamonds) and iron nuclei (blue circles) of $10^{17}$~eV (EPOS LHC model) is given.}
\label{fig-3}
\end{figure}

Finally we investigated the possibility of using $R_0$ together with $X_{\rm max}$ for two-component event-by-event analysis available in case of hybrid measurements by the  surface array and fluorescent telescopes. In Figure~\ref{fig-3} correlations and fluctuations of $R_0$ and $s$ at sea level are shown for $10^{17}$~eV vertical proton- and iron-induced showers simulated with EPOS LHC model. Strong anticorrelation useful for primary mass discrimination in individual showers is observed.

\section{Discussion and conclusions}

Discrepancies between mass composition estimates, obtained by different methods from  different air shower arrays data are probably caused by the sum of instrumental and methodical systematic biases, as well as strong model dependence of observables using as primary mass indicators (mostly in case of muon component characteristics), inability of effective discrimination of electromagnetic and muon contributions to the ground EAS signal, lack of understanding of meteorological effects and specific detectors properties etc.

A possible solution to the problem might be achieved with refined multi-hybrid measurements by existing EAS arrays taking into account the capabilities of their anticipated upgrades (e.g., separate detection of electromagnetic and muon contributions to the ground signal by \glqq Auger Prime\grqq~\cite{AugerPrime}) together with generalizations of the analysis by revealing universal features, new parameters and functionals that exhibit weak sensitivity to the interaction model being good primary mass indicators.

The slope of lateral distribution of EAS charged particles far from the shower core is known to be a primary mass discriminator. But it is rarely used in recent experimental works because precision measurements of the shape of lateral distribution can not be realised by air shower arrays with large separation between ground stations. From the other hand the important advantage of the LDF as a source of information about mass composition is that ground detectors of charged particles provides an order of magnitude higher duty cycle in comparison with atmospheric telescopes. It is important to note here that using characteristics of muon component as basic composition estimates faces the \glqq muon excess problem\grqq~\cite{muonexcess1, muonexcess2, muonexcess3}, which requires extended investigations.

On the basis of the analysis of CORSIKA simulations we show that, when adequate model-independent description of lateral distribution of electrons in the framework of scaling formalism is adopted, the single integral parameter of lateral distribution --- scaling factor $R_0$ --- is a potentially effective primary mass estimator. When combining with the data on longitudinal shower age, it provides improved composition results both for averaged shower measurements and on event-by-event basis.

The proposed approach exhibits promises as a tool for both primary composition studies and hadronic interaction models tests in wide primary energy range. It could be implemented for the present and future (multi-)hybrid EAS observations, especially in realising the potential of Auger and Telescope Array observatories upgrades, as well as for reanalysis and cross-calibration of the data collected from different air shower arrays.

This work was supported by Russian Foundation for Basic Research (grant \#16-02-01103~a).

\end{document}